\documentclass[10pt]{article}
\usepackage[OE]{express}

\usepackage{dsfont}
\usepackage{cite}
\usepackage{accents}
\usepackage{algorithm}
\usepackage{algorithmic}
\usepackage{flushend}

\usepackage{xcolor}
\definecolor{darkblue}{RGB}{0,0,128}
\definecolor{darkred}{RGB}{128,0,0}
\definecolor{darkbrown}{rgb}{0.375,0.25,0.125}

\DeclareMathOperator{\nft}{NFT}
\DeclareMathOperator{\inft}{INFT}
\DeclareMathOperator{\shift}{shift}

\newcommand{\Complex}{\mathbb{C}}

\providecommand{\norm}[1]{\bigl\lVert#1\bigr\rVert}

\def \Im {\mathop {\rm Im}}
\def \Re {\mathop {\rm Re}}

\newcommand{\abs}[1]{\ensuremath{\left|#1\right|}}

\newcommand\Let{\mathrel{\mathop:\!\!=}}

\newcommand{\vc}[1]{\ensuremath{\mathbf{#1}}}

\newcommand{\av}[1]{\ensuremath{\langle #1\rangle}}

\renewcommand{\l}{\lambda}

\begin{document}

\thispagestyle{empty}

\title{Polarization-Division Multiplexing Based on the Nonlinear Fourier Transform}

\author{Jan-Willem Goossens,\authormark{1,2,*} Mansoor I. Yousefi,\authormark{2} Yves Jaou\"en\authormark{2} and Hartmut Hafermann\authormark{1}}
\address{\authormark{1} Mathematical and Algorithmic Sciences Lab, Paris Research Center, Huawei Technologies France\\
\authormark{2} Communications and Electronics Department, Telecom ParisTech, Paris, 75013, France
}
\email{\authormark{*}jan.willem.goossens@huawei.com}

\begin{abstract}
Polarization-division multiplexed (PDM) transmission based on the nonlinear Fourier transform (NFT) is proposed for optical 
fiber communication.
The NFT algorithms are generalized from the scalar nonlinear Schr\"odinger equation for one polarization
to the Manakov system for two polarizations. The transmission performance of the PDM 
nonlinear frequency-division multiplexing (NFDM)  and PDM orthogonal frequency-division 
multiplexing (OFDM) are determined. It is shown that the transmission performance in terms of Q-factor is approximately the same in PDM-NFDM and single polarization NFDM at twice the data rate and that the polarization-mode dispersion does not seriously degrade system performance. Compared with PDM-OFDM, PDM-NFDM achieves a Q-factor gain of 6.4 dB. The theory can be generalized to multi-mode fibers in the strong coupling regime, paving the way for the application of the NFT to address the nonlinear effects in space-division multiplexing.
\end{abstract}


\bibliographystyle{osajnl}
\bibliography{Manakov}

\section{Introduction}

Nonlinear frequency-division multiplexing (NFDM) is an elegant method to address the nonlinear effects 
in optical fiber communication. The scheme can be viewed as a generalization of communication using fiber solitons~\cite{Hasegawa1973} and goes back to the original idea of eigenvalue communication~\cite{Hasegawa1993}.

In this approach, information is encoded in the nonlinear spectrum of the signal, defined 
by means of the nonlinear Fourier transform (NFT),~\cite{Gardner1967, Yousefi2014,Yousefi2014a}.
The evolution of the nonlinear spectral components in fiber  
is governed by simple independent equations.
As a result, the combined effects of the dispersion and nonlinearity can be compensated in the digital domain by the application of an all-pass-like filter. Furthermore, interference-free communication can be achieved in network environments.  

The nonlinear spectrum consists of a continuous part and, in the case of the anomalous 
dispersion fiber, also a discrete part (solitonic component). In principle all degrees-of-freedom can be modulated. Prior work 
has mostly focused on either continuous spectrum modulation~\cite{Yousefi2014b,Le2014,Le2015,Yousefi2016,Tavakkolnia2017}, or discrete spectrum modulation~\cite{Yousefi2014b}. Transmission based on NFT has been experimentally demonstrated 
for the continuous spectrum~\cite{Le2016a,Le2016c,Le2017}, discrete spectrum~\cite{Aref2015}, as well as 
the full spectrum~\cite{Aref2016}.
 
Modern coherent optical fiber systems are based on polarization-division multiplexed (PDM) transmission to improve 
the achievable rates. However, research on data transmission using the NFT is limited to 
the scalar nonlinear Schr\"odinger equation, which does not take into account polarization 
effects (with the exception of~\cite{Maruta2015}). 

In this paper we overcome this limitation by generalizing the NFDM to the Manakov system, proposing 
PDM-NFDM. We develop a stable and accurate algorithm to compute the forward and inverse NFT of a two-dimensional signal.  
It is shown that the PDM-NFDM based on the continuous spectrum modulation is feasible, and that the data rate 
can be approximately doubled compared to the single polarization NFDM. Compared to the PDM-OFDM, the PDM-NFDM
exhibits a peak Q-factor gain of 6.4 dB, in a system with 25 spans of 80 km standard single-mode fiber and 16 QAM.

In this paper, we set the discrete spectrum to zero and modulate only the
continuous spectrum. Even though not all available degrees of freedom are modulated, the achievable rates reach remarkably close to the upper bound~\cite{Yangzhang2017,Yousefi2015}.

The differential group delay and randomly varying birefringence give rise to temporal pulse broadening due to  polarization-mode dispersion (PMD). The PMD might be compensated in conventional systems, and may even be beneficial in reducing the nonlinear distortions~\cite{Serena2009}. On the other hand, PMD changes the nonlinear interaction between signals in the two polarizations. The impact of PMD on NFDM is not fully investigated yet~\cite{Tavakkolnia2017a}. In this paper, we show that linear polarization effects can be equalized in PDM-NFDM at the receiver using standard techniques, and that the PMD does not seriously degrade performance.

\section{Channel Model with Two Polarizations}

Light propagation in two polarizations in optical fiber is modeled by the 
coupled nonlinear Schr\"odinger equation (CNLSE)~\cite{Menyuk1989}. The fiber 
birefringence usually varies rapidly and randomly along the fiber in practical systems 
(on a scale of 0.3 to 100 meters). Under this assumption, the averaging of the nonlinearity in the CNLSE leads to the Manakov-PMD equation~\cite{Wai1996,Marcuse1997,Menyuk2006}:
\begin{align}
\frac{\partial \vc{A}}{\partial Z} =& -j\frac{\Delta\beta_0}{2} \sigma_{Z}\vc{A}-\frac{\Delta\beta_1}{2} \sigma_Z
\frac{\partial \vc{A}}{\partial T}\notag\\
& -\frac{\alpha}{2}\vc{A} + j\frac{\beta_2}{2}\frac{\partial^2 \vc{A}}{\partial T^2} 
- j\gamma\frac{8}{9}\norm{\vc{A}}^2\vc{A}.
\label{eq:manakov-pmd}
\end{align}
Here $\vc{A}\equiv\vc{A}(Z,T)$ is the $2\times 1$ Jones vector containing the complex envelopes $A_1$ and $A_2$ of the two polarization components, $Z$ denotes the distance along the fiber, $T$ represents time, $\sigma_Z$ is a $2\times 2$ Pauli matrix (depending 
on the state of polarization at $Z$), and 
$\Delta\beta_0$, $\Delta\beta_1$, $\alpha$, $\beta_2$ and $\gamma$ are constant numbers. 

The term $\partial\vc{A}/\partial T$ is responsible for the PMD~\cite{Menyuk2006}, while the second line represents loss, chromatic dispersion and Kerr nonlinearity. The factor $8/9$ in front of the nonlinear term stems from polarization averaging. 

The NFT applies to the integrable equations. However, the Manakov-PMD system \eqref{eq:manakov-pmd} 
apparently is not integrable in the presence of loss or PMD. The loss may be approximately compensated 
using ideal distributed Raman amplification, leading to a lossless model  with distributed
additive white Gaussian noise (AWGN). Discrete amplification with short spans can also be modeled similarly, but with a modified nonlinearity coefficient (depending on loss); see \cite{Le2015}. 
Ignoring the PMD and, for the moment, noise, \eqref{eq:manakov-pmd} is reduced to 
\begin{align}
\frac{\partial \vc{A}}{\partial Z} =& j\frac{\beta_2}{2}\frac{\partial^2 \vc{A}}{\partial T^2} 
- j\gamma\frac{8}{9}\norm{\vc{A}}^2\vc{A}.
\label{eq:manakov_int}
\end{align}

It is convenient to normalize \eqref{eq:manakov_int}. Let $Z_0=1$ and  
\begin{equation}
T_0=\sqrt{\bigl(\abs{\beta_2}Z_0\bigr)/2},\quad 
A_0=\sqrt{2\Bigl/\bigl(\frac{8}{9}\gamma Z_0\bigr)}.
\label{eq:normalizationManakov}
\end{equation}
Introducing the normalized variables $z=Z/Z_0$, $t=T/T_0$ and $\vc{q}=\vc{A}/A_0$, \eqref{eq:manakov_int}
is simplified to the Manakov equation
 \begin{align}
j\frac{\partial \vc{q}}{\partial z} =& \frac{\partial^2 \vc{q}}{\partial t^2} 
- 2s \norm{\vc{q}}^2\vc{q},
\label{eq:manakov_norm2}
\end{align}
where $s=1$ in the normal dispersion (defocusing) regime and $s=-1$ in the anomalous dispersion (focusing or solitonic) regime. 

It is shown in ~\cite{Ablowitz1981} that \eqref{eq:manakov_norm2} is integrable. In what follows, we develop the 
corresponding NFT.

\section{NFT of the Two-Dimensional Signals}

In this section, basics of the NFT theory of the two-dimensional signals, as well as numerical algorithms to 
compute the forward and inverse NFT, are briefly presented \cite{Manakov1974}. 

\subsection{Brief Review of the Theory}

Equation \eqref{eq:manakov_norm2} can be represented by a Lax pair $\hat{L}$ and $\hat{M}$. This means that, 
operators $\hat{L}$ and $\hat{M}$ can be found such that \eqref{eq:manakov_norm2} is in one-to-one correspondence with
the Lax equation $\partial\hat{L}/\partial z = [\hat{M},\hat{L}]$. The Lax pair for the Manakov equation 
was found by Manakov in 1974~\cite{Manakov1974,Degasperis2016}. The operator $\hat{L}$ is:
\begin{equation}
\hat{L} = j\begin{pmatrix}
\frac{\partial }{\partial t} & -q_1 & -q_2\\
sq_1^* &  -\frac{\partial }{\partial t} & 0\\
sq_2^* &0&-\frac{\partial }{\partial t}
\end{pmatrix}.
\label{eq:defL}
\end{equation} 
To simplify the presentation, consider the focusing regime with $s=-1$.   

The eigenvalue problem 
\begin{equation}
\hat L v=\lambda v 
\label{eq:eig-prob}
\end{equation}
can be solved for the Jost function $v$ (eigenvector) assuming that the 
signals vanish at $t=\pm\infty$. This gives rise to six boundary conditions for $v$, denoted by $j_{\pm}$:  
\begin{eqnarray}
j_\pm^{(0)}\rightarrow e^{(0)}\exp(-j\lambda t), \quad j_\pm^{(i)}\rightarrow e^{(i)}\exp(j\lambda t), \quad i=1,2,
\quad
\textnormal{as} \quad t\rightarrow \pm\infty,
\label{eq:boundaries}
\end{eqnarray}
where $e^{(k)}$ are unit vectors, i.e., $e^{(k)}_l=\delta_{kl}$, $k,l=0,1,2$. Each of the boundary conditions 
\eqref{eq:boundaries} is bounded when $\lambda\in\Complex^+$ or $\lambda\in\Complex^-$. The eigenvalue 
problem \eqref{eq:eig-prob} under the boundary conditions \eqref{eq:boundaries} can be solved,  obtaining  
six Jost functions  $\{j^{(i)}_\pm(t,\lambda)\}_{i=0,1,2}$ 
for all $t$. It can be shown that $\{j^{(i)}_{+}(t,\l)\}_{i=0,1,2}$ and $\{j^{(i)}_{-}(t,\l)\}_{i=0,1,2}$ each form an 
orthonormal basis for the solution space of \eqref{eq:eig-prob}.
Thus we can expand $j^{(0)}_-(t,\lambda)$ in the basis of $\{j^{(i)}_{+}(t,\l)\}_{i=0,1,2}$:
\begin{equation}
j_-^{(0)}(t,\l) = a(\l) j_+^{(0)}(t,\l)+b_1(\l) j_+^{(1)}(t,\l)+b_2(\l) j_+^{(2)}(t,\l),
\label{eq:jostexp}
\end{equation}
where $a(\l)$, $b_1(\l)$ and $b_2(\l)$ are called nonlinear Fourier coefficients. It can be shown 
that in the focusing regime the inner product of two Jost functions corresponding to the same eigenvalue does not depend on time. This implies that $a(\lambda)$ and $b_i(\lambda)$ do not depend on time (as the notation in \eqref{eq:jostexp} suggests). 
The NFT of the $\vc{q}=[q_1,q_2]$ is now defined as
\begin{align}
\text{NFT}(\vc{q})(\l) = &\begin{cases}
\hat{q}_i = \frac{b_i(\l)}{a(\l)}, & \lambda\in\mathds{R},\vspace{1mm}\\ \tilde{q}_i = \frac{b_i(\l_j)}{a'(\l_j)}, & \l_j \in \mathds{C}^+,
\end{cases}
\quad i=1,2,
\label{eq:nft}
\end{align}
where $\l_j$, $j=1,2,\cdots, N$, are the solutions of $a(\l_j)=0$ in $\l_j\in\mathds{C}^+$.

An important property of $\text{NFT}(q(t,z))(\l)$ is that it evolves in distance according to the all-pass-like filter
\begin{align}
H(\lambda) = e^{4js\lambda^2 L}.
\label{eq:nlphase}
\end{align}

\emph{Remark.}
Note that if $q_1$ or $q_2$ is set to zero in the Manakov equation and the associated $\hat{L}$ operator \eqref{eq:defL}, the
equation and the operator are reduced, respectively, to the scalar NLSE and the corresponding $\hat{L}$ operator~\cite{Yangzhang2017}. 
Also, the theory in this section can straightforwardly be generalized to include any number of 
signals $q_i$, $i=1,2,...$, for instance, LP modes propagating in a multi-mode fiber in the strong coupling regime.

\emph{Remark}
Since the Jost vectors are orthonormal at all times, \eqref{eq:jostexp} implies the unimodularity condition
\begin{equation}
|a(\lambda)|^2+|b_1(\lambda)|^2+|b_2(\lambda)|^2 = 1,
\label{eq:unimod}
\end{equation}
which will be used in the algorithm.

\subsection{Forward NFT Algorithm}

We develop the NFT algorithms for the continuous spectrum that is considered in this paper. The forward and inverse NFT algorithms are, respectively, based on the Ablowitz-Ladik and discrete layer peeling (DLP) methods. These algorithms generalize the corresponding ones in~\cite{Yousefi2016}.

We begin by rewriting the eigenvalue problem $\hat{L}v=\lambda v$ in the form $\partial v/\partial t = Pv$, where 
\begin{equation}
P = \begin{pmatrix}
	-j\lambda&q_1(t)&q_2(t)\\
	-q_1^*(t)&j\lambda&0\\
	-q_2^*(t)&0&j\lambda
\end{pmatrix}.
\end{equation}
Here and in the remainder of this section we suppress the dependence on the coordinate $z$.
We discretize the time interval $[T_0,T_1]$ according to $t[k] = T_0 + k\Delta T$, where $\Delta T= (T_1-T_0)/N$ such that $t[0]=T_0$ and $t[N]=T_1$. We set $q_i[k] = q_i(T_0+k\Delta T)$ and similarly for the vector $v$. The Ablowitz-Ladik method is a
discretization of $\partial v/\partial t = Pv$ as follows 
\begin{equation}
v[k+1,\l] = c_k\begin{pmatrix}
	z^{1/2}&Q_1[k]&Q_2[k]\\-Q_1[k]^*&z^{-1/2}&0\\-Q_2[k]^*&0&z^{-1/2}
\end{pmatrix}v[k,\l],
\label{eq:forwardit}
\end{equation}
where   $Q_i[k]=q_i[k]\Delta T$, $z:= e^{-2j\lambda\Delta T}$ and $k=0,1,\cdots, N-1$.  With this discretization $v$ becomes a periodic function of $\l$ with period $\pi/\Delta T$. We introduced a normalization factor $c_k=1/\sqrt{\abs{\det P[k]}}$ in \eqref{eq:forwardit}, where 
$\abs{\det P[k]}=1+\abs{Q_2[k]}^2+\abs{Q_2[k]}^2$, to improve numerical stability. 

The iterative equation \eqref{eq:forwardit} is initialized with 
\begin{equation*}
v[0,\l]=j_-^{(0)}(T_0,\l)=e^{(0)}z^{T_0/2\Delta T}.
\end{equation*}
After $N$ iterations, the nonlinear Fourier coefficients are obtained as projections onto the Jost-solutions $j_+^{(i)}(T_1,\lambda)$,
\begin{align}
a[\l] &= z^{-\frac{N}{2}-\frac{T_0}{2\Delta T}}v_0[N,\l],\\
b_{i}[\l] &= z^{\frac{N}{2}+\frac{T_0}{2\Delta T}}v_{i}[N,\l],\quad i=1,2.
\end{align}
The continuous component in the $\nft(q)(\lambda)$ is then computed based on \eqref{eq:nft}.

The algorithm can be efficiently implemented in the frequency domain. 
Let us write~\eqref{eq:forwardit} as
\begin{equation}
V[k+1,\l] = c_k\begin{pmatrix}
	1 &Q_1[k]z^{-1}&Q_2[k]z^{-1}\\-Q_1^*[k]&z^{-1}&0\\-Q_2^*[k]&0&z^{-1}
\end{pmatrix}V[k,\l],
\label{eq:AL-2}
\end{equation}
where $V[k,\l]\Let (A[k,\l],B_1[k,\l],B_2[k,\l])^T$, and 
\begin{align}
A[k,\lambda] &= a[k,\lambda]\nonumber\\
B_{i}[k,\lambda] &=z^{-N-\frac{T_0}{\Delta T}+\frac{1}{2}}b_{i}[k,\l],\quad i=1,2.
\label{eq:Afromv}
\end{align}
We discretize $\lambda$ on the interval $[\Lambda_0,\Lambda_1]$ with $\Lambda_1=-\Lambda_0=\pi/(2\Delta T)$ such that $\lambda[k] = \Lambda_0+k\Delta \Lambda$. Let tilde '$\sim$' denote the action of the discrete Fourier transform DFT with respect to $\l[k]$, e.g.,
$\tilde{A}[.,l]=\textnormal{DFT}(A[.,\l[k]])$, where $l=0,1,,\cdots,N-1$ is the discrete frequency. Note that 
$\textnormal{DFT}\bigl\{z^{-1}B_i[.,\l[k]]\bigr\}[l]=\shift\big\{\tilde{B_i}\big\}[l]$, where $\shift$ denotes circular right shift of the array by one element.  Equation \eqref{eq:AL-2} in the frequency domain is
\begin{align}
\tilde{A}[k+1,l] &= c_k(\tilde{A}[k,l]+Q_1[k]\shift\big[\tilde{B_1}[k]\big][l]\notag\\
&\qquad\qquad\enspace\enspace +Q_2[k]\shift\big[\tilde{B_2}[k]\big][l]),\notag\\
\tilde{B_1}[k+1,l] &= c_k(	-Q_1[k]^*\tilde{A}[k,l]+\shift\big[\tilde{B_1}[k]\big][l]),\label{eq:fftforwardit}\\
\tilde{B_2}[k+1,l] &= c_k(    -Q_2[k]^*\tilde{A}[k,l]+\shift\big[\tilde{B_2}[k]\big][l]).\notag
\end{align}
The initial condition is given by $\tilde{A}[0] = \text{DFT}\big[a[k]\big][0]$ and $\tilde{B_i}[0] = 0$. At $k=N-1$, $a$ and $b_i$ are 
found by recovering $V[N,\lambda]$ through an inverse DFT and using~\eqref{eq:Afromv}.

\subsection{Inverse NFT Algorithm}

The inverse NFT algorithm consists of two steps. First, we compute $v[N,\lambda]$ from the continuous spectra
$\hat{q}_{1,2}(\lambda)$ and invert the forward iterations~\eqref{eq:forwardit}. Second, at each iteration, 
we compute $Q[k]$ from $v[k,\lambda]$.

Substituting $b_i(\l)=\hat{q}_i(\l) a(\l)$ in the unimodularity condition~\eqref{eq:unimod}, we can compute $|a(\l)|$ 
\begin{equation*}
|a(\l)| = 1/\sqrt{1+|\hat{q}_1(\l)|^2+|\hat{q}_2(\l)|^2}.
\end{equation*}
In the absence of a discrete spectrum, from~\eqref{eq:nft}, $a(\l)\neq 0$ for all $\l$. We have $\Re(\log(a)) = \log(|a|)$ and $\Im(\log(a)) = \angle a$. Since $a(\l)$ and therefore $\log(a)$ are analytic functions of $\l$ we can recover the phase of $a$ as $\angle{a} = \mathcal{H}(\log(|a|))$, where $\mathcal{H}$ denotes the Hilbert transform.
 
The inverse iterations are obtained by inverting the matrix in~\eqref{eq:forwardit} and dropping 
terms of order $Q^2_i\sim\Delta T^2$ (to yield the same accuracy as for the forward iterations):
\begin{equation}
v[k,\l] = c_k\begin{pmatrix}
	z^{-1/2}&-Q_1[k]&-Q_2[k]\\Q_1^*[k]&z^{1/2}&0\\Q_2^*[k]&0&z^{1/2}
\end{pmatrix}v[k+1,\l].
\end{equation}

In practice, as in the forward NFT, it is better to invert the frequency domain iterations \eqref{eq:fftforwardit}.
The signal $Q[k]$ can then be obtained from $V[k+1,l]$ as follows. Recall that 
$\shift\big[\tilde{B}_i[k]\big][l]=\tilde{B}_i[k,l-1]$. Using the initial conditions for the \emph{forward} iterations $\tilde{A}[0,l]=\delta_{0,l}$ and $\tilde{B}_i[0,l]=0$, it is straightforward to show 
that $\tilde{A}[k,l]=0$ and $\tilde{B}_i[k,l]=0$ for $l \geq k>0$. In particular, $\tilde{B}_i[k,N-1]=0$ for $k<N$.
For the first element of $\tilde{B}_i[k]$, from~\eqref{eq:fftforwardit} and $\shift\big[\tilde{B_1}[k]\big][-1] = \shift\big[\tilde{B_1}[k]\big][N-1]=0$, we obtain
\begin{align*}
\tilde{A}[k+1,0] &= c_k\tilde{A}[k,0],\\
\tilde{B_1}[k+1,0] &= -c_kQ_1[k]^*\tilde{A}[k,0],\\
\tilde{B_2}[k+1,0] &= -c_kQ_2[k]^*\tilde{A}[k,0].
\end{align*}
These equations can be solved for $Q_i^*[k] = -\tilde{B}_{i}[k+1,0]/\tilde{A}[k+1,0]$.

\subsection{Testing the NFT Algorithms}
\label{sec:nfttest}

\begin{figure}%
\includegraphics[width=0.5\columnwidth]{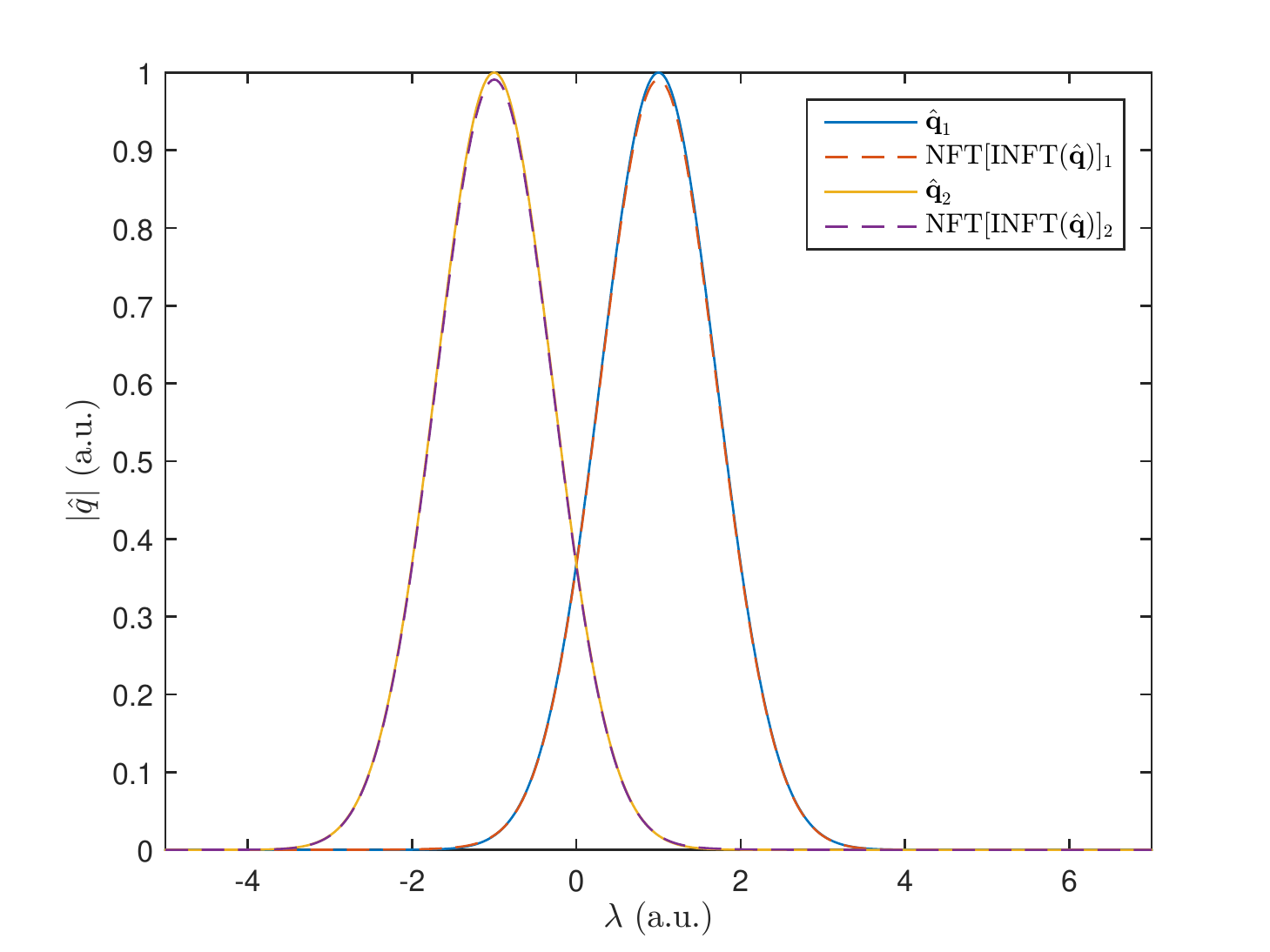}%
\includegraphics[width=0.5\columnwidth]{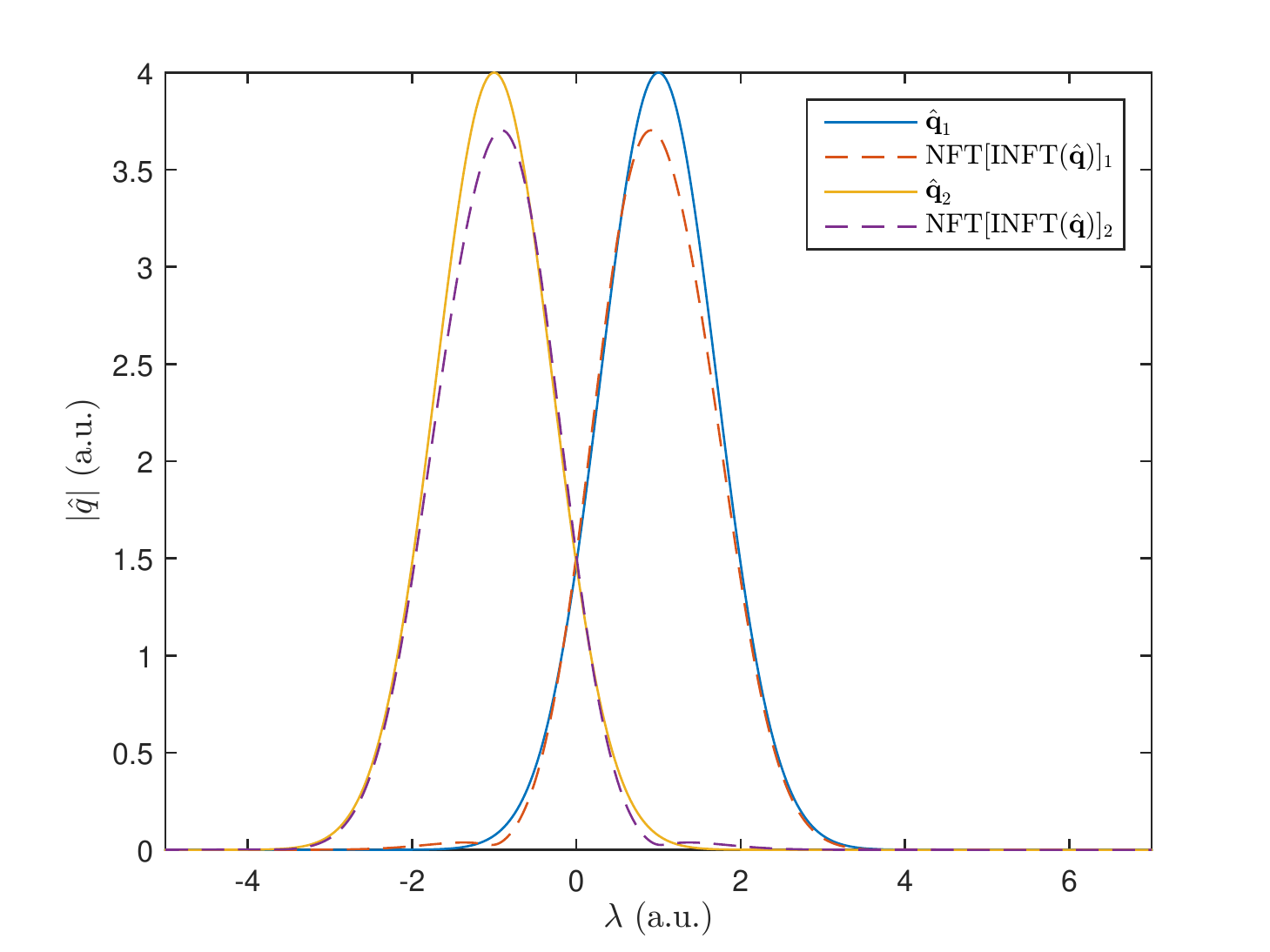}%
\caption{Comparing $\hat{\vc{q}}=[\hat{q}_1,\hat{q}_2]$ and  $\nft(\inft(\hat{\vc{q}}))$, where $\hat{q}_1$ and $\hat{q}_2$  are displaced Gaussians. For fixed sampling rate, the algorithm is less accurate at higher input-power.}%
\label{fig:proofofprinciple}%
\end{figure}

The forward and inverse NFT numerical algorithms are tested as follows. 
Figure~\ref{fig:proofofprinciple} compares a signal containing two polarization components in the nonlinear Fourier domain with its reconstruction after successive INFT and NFT operations. We take two displaced Gaussians with standard deviation $\sigma=\sqrt{2}$ as input signals for the two polarization components $\hat{\vc{q}}_i$, $i=1,2$.
We set time windows to $T=64$ and take 2048 samples in time and nonlinear Fourier domain. In the discrete layer peeling method, the signal is periodic in nonlinear Fourier domain with period $\pi/\Delta T\approx 100$. 

Deviations occur as the amplitude of the signal is increased. The accuracy of the NFT (INFT) can be 
enhanced by increasing the sampling rate in time (nonlinear frequency), as this reduces errors 
due to to the piecewise-constant approximation of the signal. We find that we roughly need twice the 
number of samples to achieve the same accuracy as in the single-polarization case.

\section{PDM-NFDM Transmitter and Receiver}
\label{sec:system}

\begin{figure*}%
\begin{center}
\includegraphics[width=0.9\textwidth]{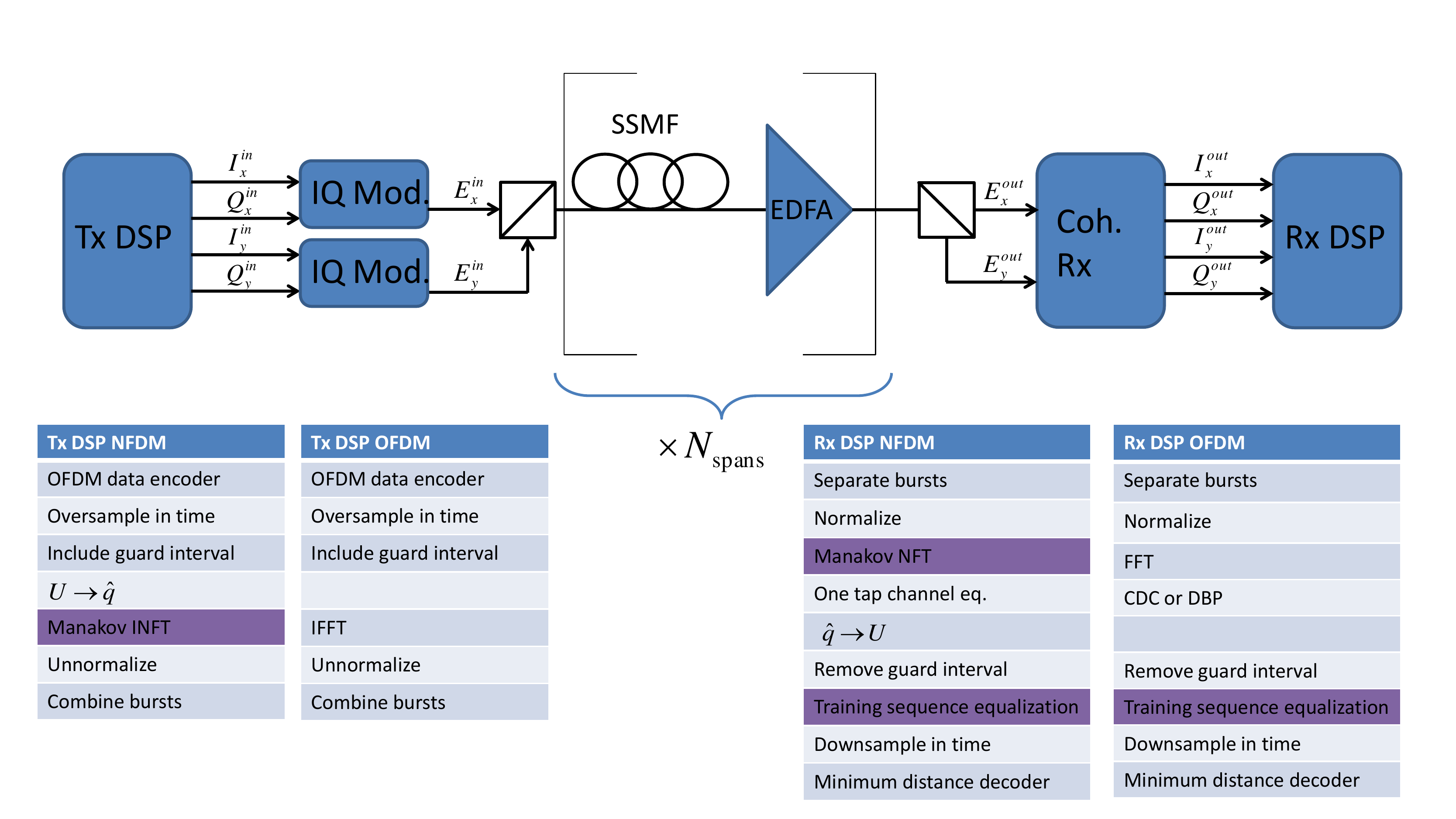}
\end{center}
\caption{System diagram for the polarization-multiplexed NFDM and OFDM transmission systems with processing steps in the transmitter and receiver DSP. Steps highlighted in purple require joint processing of both polarization components. For further explanation see text.}%
\label{fig:system}%
\end{figure*}

In this section, we describe the transmitter (TX) and receiver (RX) digital signal processing (DSP) in PDM-NFDM and PDM-OFDM. A schematic diagram of the polarization-division multiplexed NFDM and OFDM transmission systems is shown in Fig.~\ref{fig:system}. 
The TX DSP produces digital signals for the in-phase and quadrature components of both polarization components, which are fed into the IQ-Modulator. The modulated signals for the two polarization components are combined in the polarization beam combiner before they enter the transmission line consisting of multiple fiber spans. We consider the practically relevant case of lumped amplification using Erbium-doped fiber amplifiers (EDFAs). After propagation through the fiber the two polarization components are separated in the polarization beam splitter and fed into the polarization-diversity intradyne coherent receiver, which provides the input to the RX DSP.

We briefly describe the TX and RX DSP in OFDM and NFDM. We first map the incoming bits  to 
signals taken from a QAM constellation. We then oversample the discrete-time 
signal in the time domain (by introducing zeros in the frequency domain outside the support of the signal) and then 
add guard intervals in the time domain. Increasing the guard interval increases the accuracy of the INFT. Unless stated otherwise, we do not increase the guard intervals during the computation of the INFT as the amplitude is increased, so there is a penalty due to inaccuracies in the algorithm.
In NFDM, these steps are performed in what is called the $U$-domain; see~\cite{Yousefi2016}. The signal is subsequently mapped from the $U$-domain to nonlinear Fourier domain through the transformation
\begin{equation}
U_i (\lambda) = \sqrt{-\log (1-\abs{\hat{q}_i(\lambda)}^2)}e^{j\angle \hat{q}_i(\lambda)},\quad i=1,2.
\label{eq:udomain}
\end{equation}
This step has no analogue in OFDM. Note that contrary to the single-polarization case here the energy of the signal in the time  domain is only approximately proportional to the energy in the $U$-domain. At this point, we perform the inverse NFT in the case of NFDM and an inverse FFT in case of OFDM. Then we obtain the unnormalized 
signal by introducing units using \eqref{eq:normalizationManakov}. Finally we combine the OFDM or NFDM bursts to obtain the signal to be transmitted. The output of the DSP are the in-phase and quadrature components of the signals in the two polarization components. All steps in the DSP are performed independently for the two polarizations, except for the INFT, which requires joint processing.

At the RX we invert the steps of the TX DSP. The signal processing begins with burst separation and signal normalization using \eqref{eq:normalizationManakov}, followed by the forward NFT (FFT) in case of NFDM (OFDM). In NFDM, we subsequently equalize the
channel using \eqref{eq:nlphase}, which is  a single-tap phase compensation. Similarly, in OFDM we 
either compensate the dispersion with the phase $\exp(-j\beta_2\omega^2 N_\text{span}L/2)$, or perform digital 
backpropagation (DBP) for a fixed number of steps per span.
We then remove the guard intervals in time, down-sample the signal in the time domain, and obtain the output 
symbols.  The bit error rate (BER) is calculated using a minimum distance decoder for symbols.

\begin{figure*}%
\begin{minipage}{\textwidth}
\begin{center}
\includegraphics[width=0.45\textwidth]{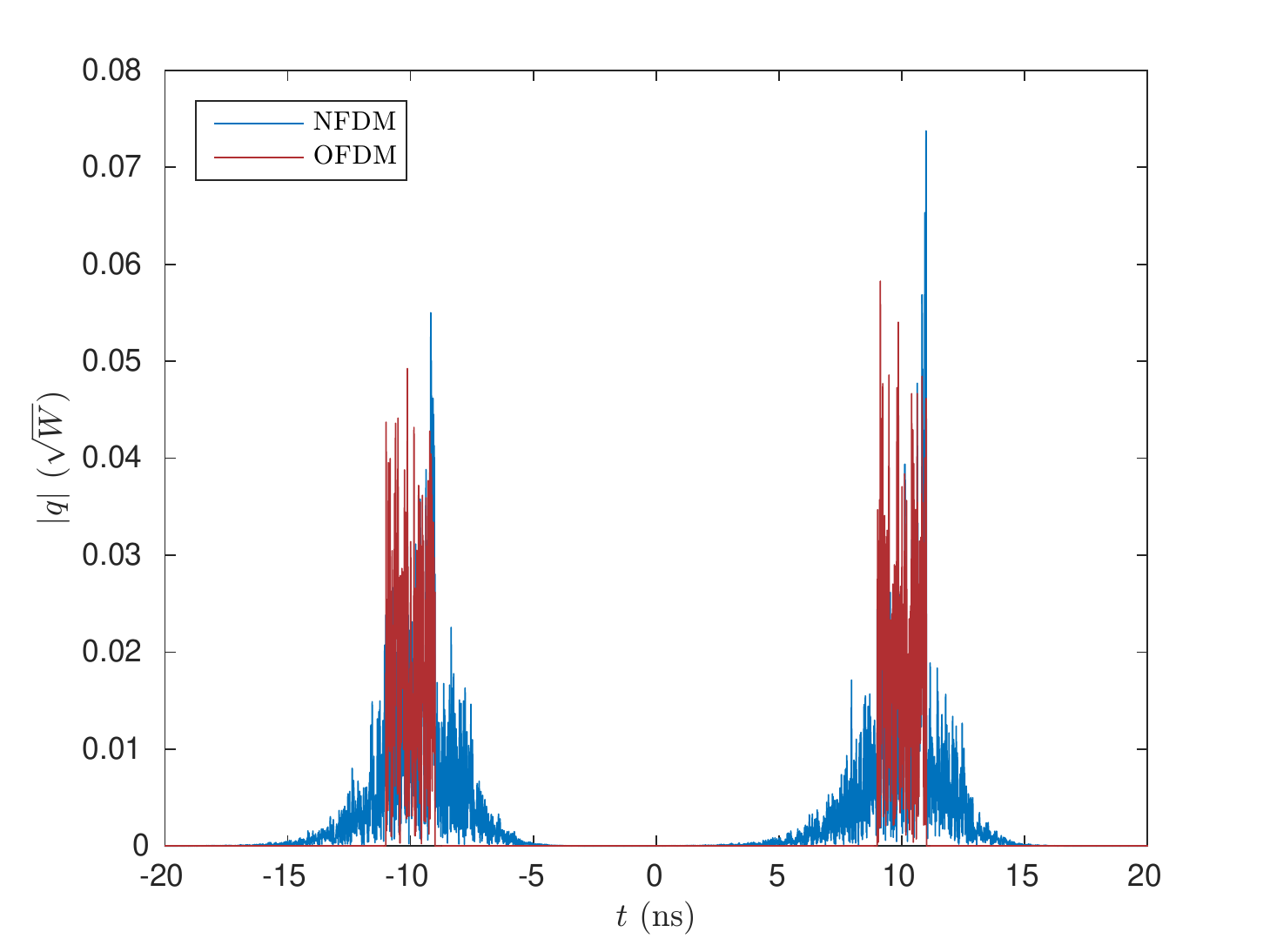}
\includegraphics[width=0.45\textwidth]{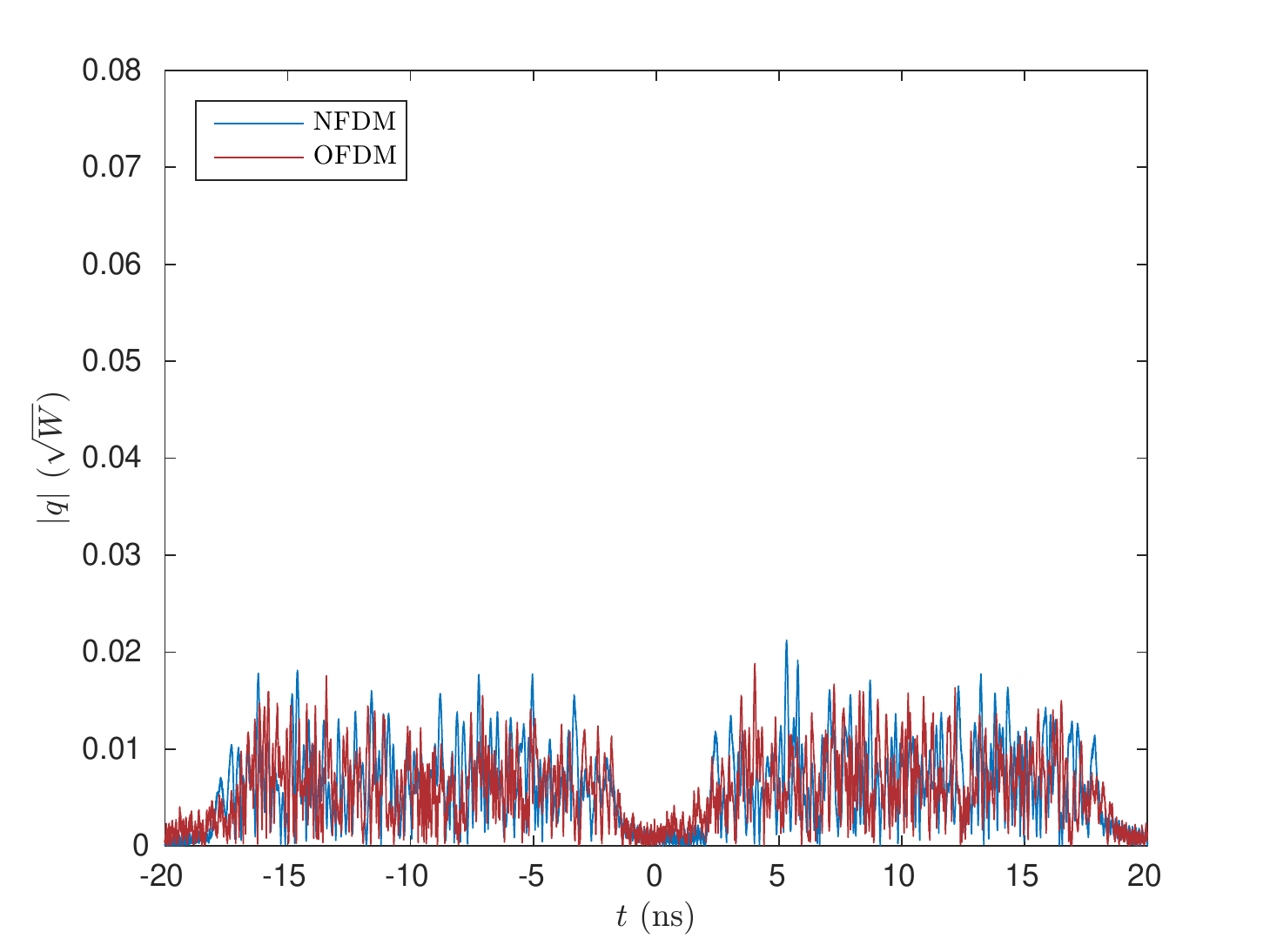}
\end{center}
\end{minipage}
\caption{NFDM and OFDM signals at the TX (left panel) and RX (right panel) using the ideal model~(\ref{eq:manakov_int}) 
at $P=0.5$ dBm without noise. Only one of the two polarizations is shown at both TX and RX. The NFDM signal at the TX
is broader due to the nonlinear effects in the NFT. At the RX, the time duration of both signals is about the same.}%
\label{fig:timesignals}%
\end{figure*}

\begin{table}%
\begin{center}
\includegraphics[width=0.4\columnwidth]{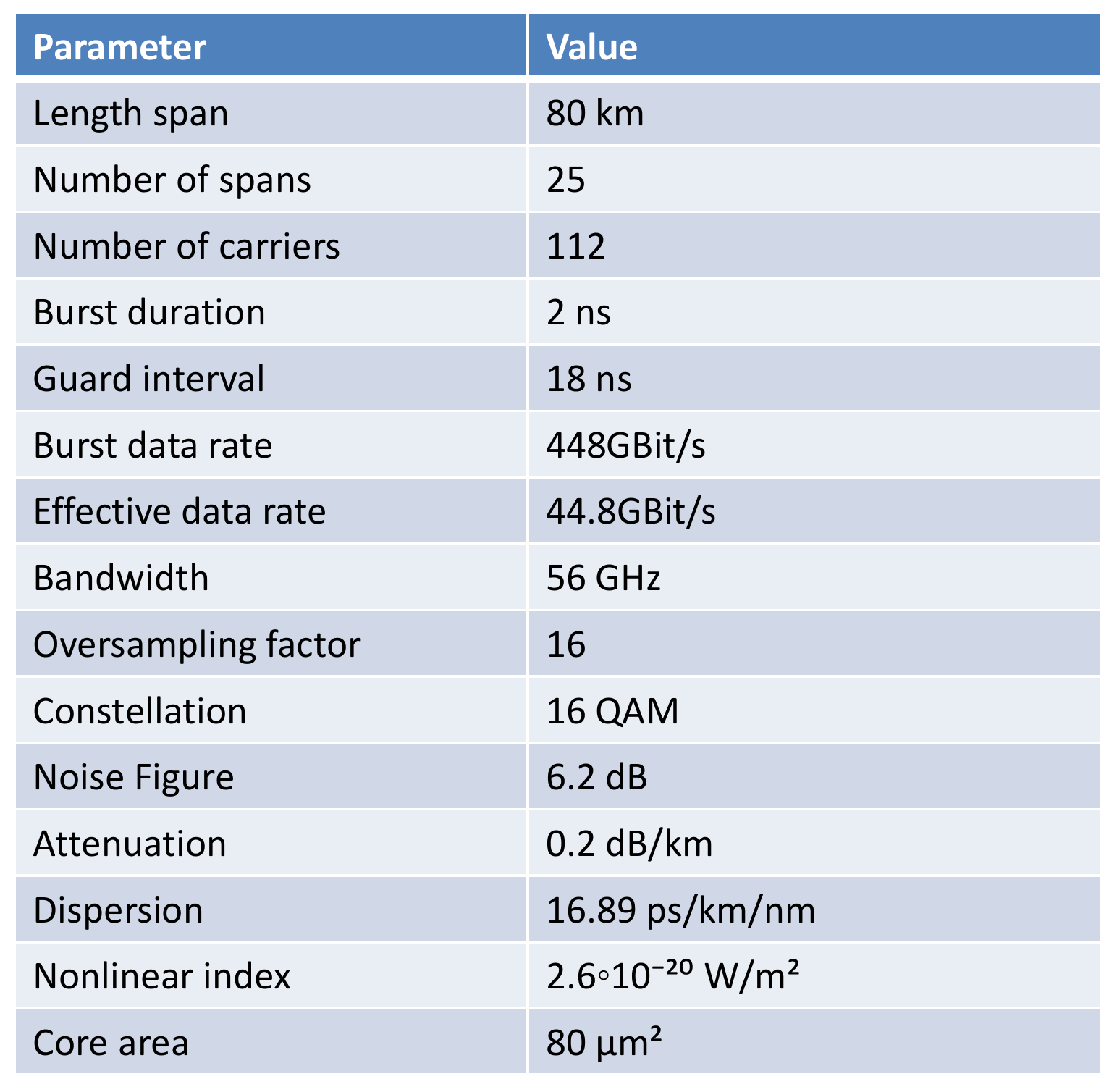}%
\end{center}
\vspace{-0.8cm}
\caption{The system parameters used in the simulations.}%
\label{tab:params}%
\end{table}

In Fig.~\ref{fig:timesignals} we compare the time domain signals of NFDM and OFDM at the TX for one of the polarization components.
We use 112 subcarriers over a burst duration of $T_0=2$ns in both cases. These parameters are the same as in Ref.~\cite{Le2015} and correspond to a burst data rate of 224GBit/s per polarization at a Baudrate of 56GBaud.
In NFDM the shape of the signal in the time domain changes as the amplitude $|\hat{q}_{1,2}|$ 
is increased in the nonlinear Fourier
domain. At a signal power of 0.5 dBm, one can see a significant amount of broadening of the NFDM signal in the time
domain. This effect makes it difficult to control the time duration of signals in NFDM.

Pulse broadening in the time domain due to the chromatic dispersion can be estimated by~\cite{Agrawal2007}
\begin{equation}
\Delta T = 2\pi |\beta_2| LB,
\label{eq:Delta-T}
\end{equation}
where $B$ is the signal bandwidth. The guard time intervals for minimizing the interaction among bursts can be approximated
using~\eqref{eq:Delta-T}. For the parameters in Table~\ref{tab:params}, we obtain $\Delta T\sim 15$ns, to which we add a 20\% margin 
to obtain a total symbol duration of $T=T_0+T_\text{guard}=(2+18)$ns$=20$ns. This corresponds to an effective data rate of 44.8Gbit/s.
The right panel of Fig.~\ref{fig:timesignals} shows the time domain signals in NFDM and OFDM in one polarization at the RX after propagation over $2000$ km governed by the integrable model~\eqref{eq:manakov_int} (we consider transmission over the realistic fiber model~\eqref{eq:manakov-pmd} below in Sec.~\ref{sec:pmd}). We observe that the amount of the temporal broadening of the 
NFDM and OFDM signals is about the same. In our simulations we do not employ pre-compensation at the TX, which may be added 
to further increase the data rate~\cite{Civelli2017}.

\section{Simulation results}

In this section, we consider the system shown in Fig.~\ref{fig:system} and compare the PDM-OFDM and 
PDM-NFDM via simulations, taking into account loss and PMD. First, we consider the model with 
loss and periodic amplification,  setting the 
PMD to zero (namely, neglecting the terms in the first line of the Eq.~\ref{eq:manakov-pmd}). Next, the PMD effects are studied 
in Sec.~\ref{sec:pmd}. The system parameters used are summarized in Table~\ref{tab:params}. 

\subsection{Effect of loss and lumped amplification}

In the lumped amplification scheme, the signal is periodically amplified after each span. The channel is described by the model~\eqref{eq:manakov-pmd} including the attenuation term, which is manifestly not integrable. The effect of lumped amplification on NFT transmission has been studied previously~\cite{Le2015}.

\begin{figure}%
\begin{center}
\includegraphics[width=0.7\columnwidth]{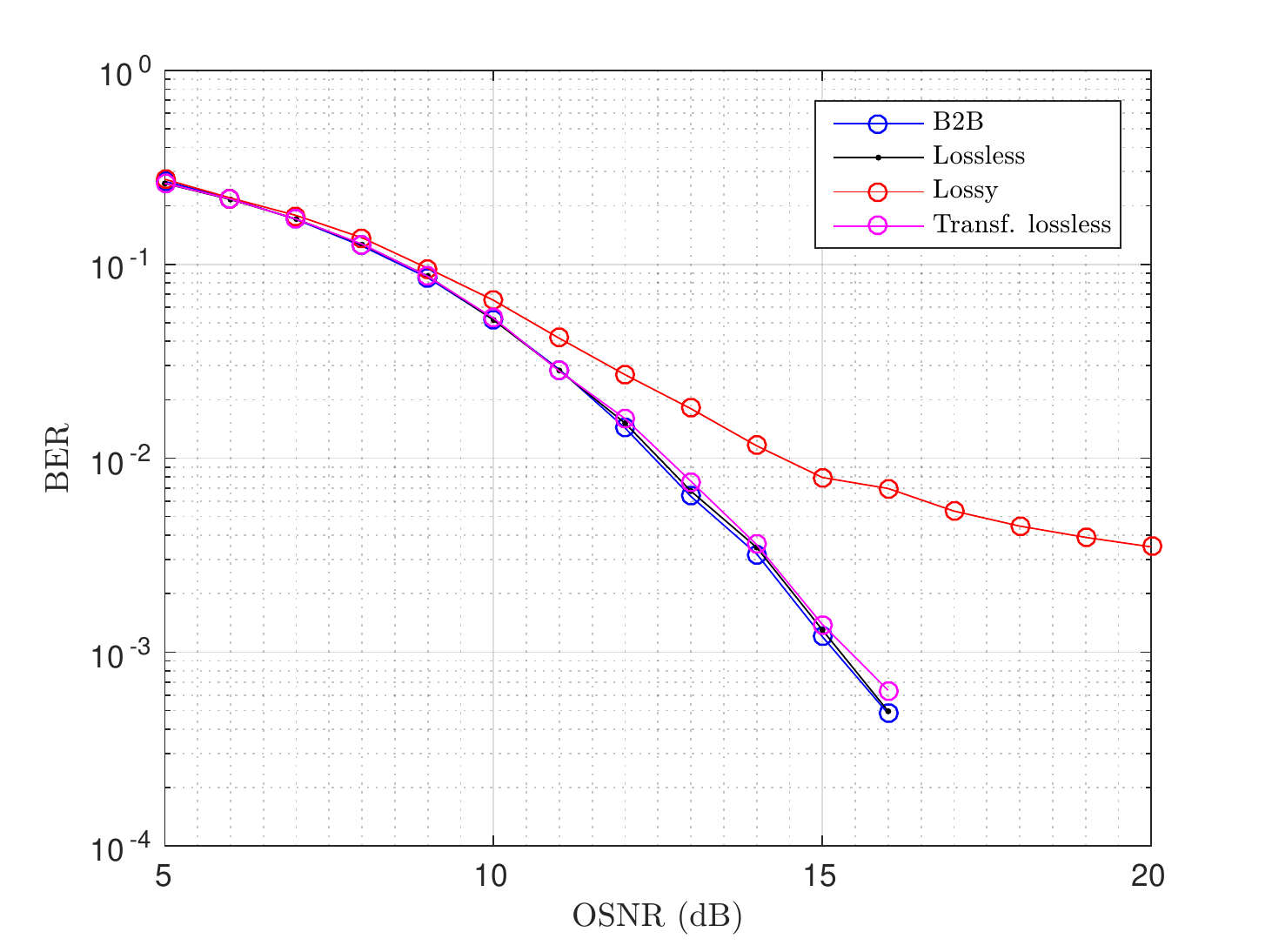}%
\caption{BER as a function of OSNR for NFDM, for back-to-back, lossless, lossy and transformed-lossless models.}
\label{fig:gammaeff}%
\end{center}
\end{figure}

We illustrate the effect of the attenuation by comparing the BER as a function of the OSNR in Fig.~\ref{fig:gammaeff} for
four models: 1) the back-to-back (B2B) configuration; 2) lossless model \eqref{eq:manakov_int} (there is
propagation but without loss); 
3) lossy model with lumped amplification; 
and 4) a transformed-lossless model that is introduced below. In producing Fig.~\ref{fig:gammaeff}, we 
artificially introduced AWGN at the receiver, to exclude the effects of the 
signal-noise interaction from the comparison. We fix the power $P=-3.1$dBm to keep the accuracy of the NFT algorithms the same and vary the OSNR by changing the noise power.

We see that the BER in B2B and lossless models are approximately the same.  This 
result verifies that the implementation of the NFT is correct. 
It also shows that the effect of burst interaction due to the finite guard interval is negligible.

The BER of the lossy model is significantly higher than the BER in the B2B and lossless models at high OSNRs. 
The BER of the lossy model seems to flatten out as the OSNR is increased. 

Fiber loss can be taken into account in the NFT as follows. The attenuation term in \eqref{eq:manakov-pmd} can be 
eliminated by a change of variable $A(z,t)=A_0(z,t)e^{-\frac{\alpha}{2}z}$~\cite{Mollenauer1991,Blow1991}. This transforms 
the non-integrable equation \eqref{eq:manakov-pmd} to the integrable equation \eqref{eq:manakov_int} (thus with zero loss) 
and a modified nonlinearity parameter 
\begin{equation}
\gamma_\textrm{eff}(L) = \frac{1}{L}\int_0^L \gamma e^{-\alpha z}dz = \gamma(1-e^{-\alpha L})/(\alpha L).
\end{equation}
We refer to the resulting model as the transformed-lossless model.

Note that the power of the signal $A_0(z,t)$ in the transformed-lossless model is higher than the power of the $A(z,t)$ in the original lossy model. That is because $\gamma_\text{eff}<\gamma$, which yields higher amplitudes according to~\eqref{eq:normalizationManakov}.

Figure~\ref{fig:gammaeff} shows that the proposed scheme for loss cancellation is indeed very effective. Using $\gamma_\text{eff}(L)$ in the inverse and forward NFTs, the BER of the lossy and transformed-lossless models 
are almost identical.

\begin{figure}%
\begin{center}
\includegraphics[width=0.7\columnwidth]{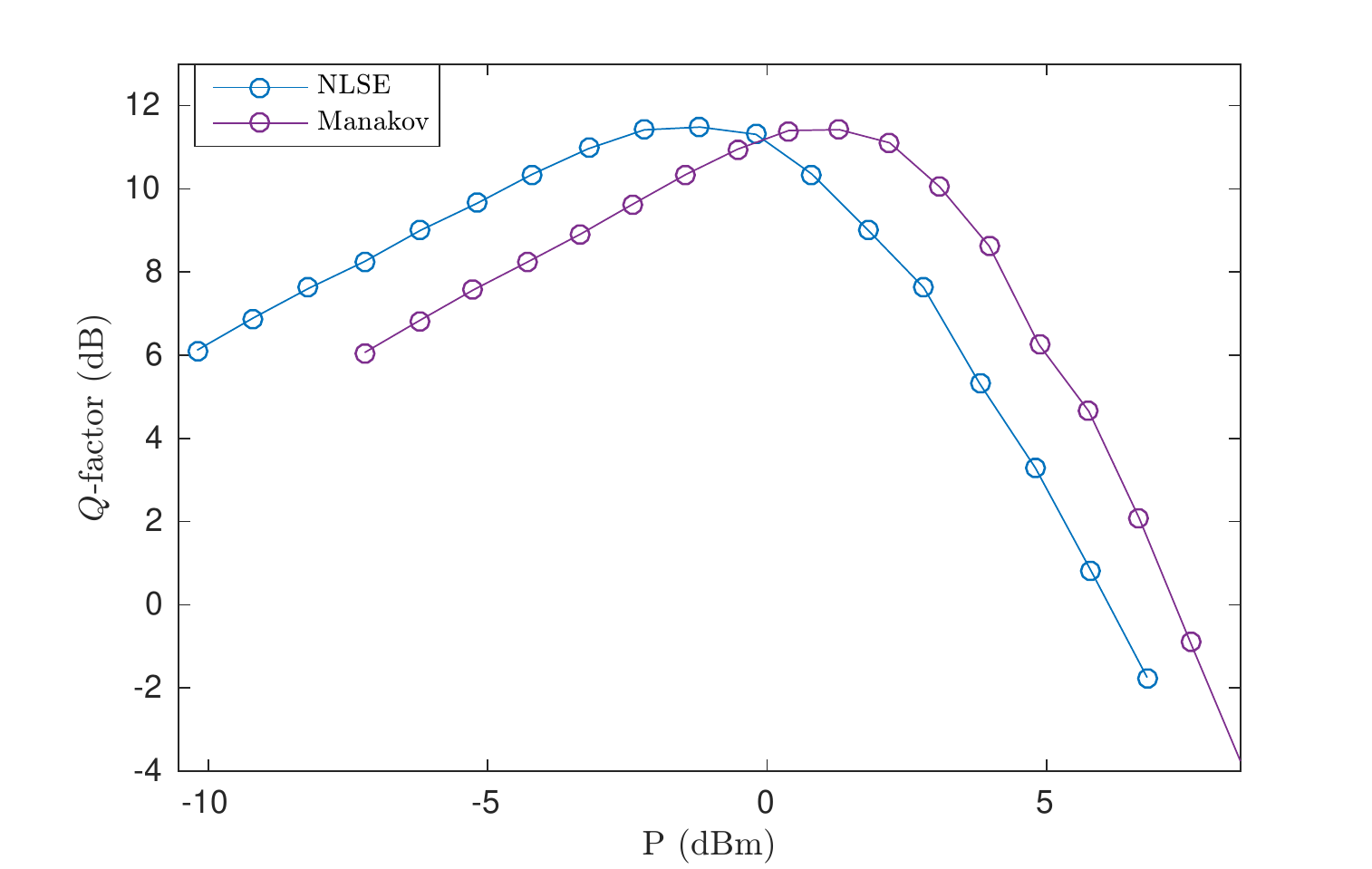}%
\caption{Comparison between NFT transmission based on the NLSE (single polarization) and the Manakov equation (polarization-multiplexed). $P$ denotes the total power of the signal. At low power the offset between the curves is 3dB, as the signal power doubles when two polarization components propagate. At higher power signal-noise interactions decrease this gap. }%
\label{fig:compare_nlse}%
\end{center}
\end{figure}

\subsection{Transmission performance PDM-NFDM}

In order to assess the transmission performance of the PDM-NFDM transmission, we performed 
system simulations as described in Sec.~\ref{sec:system}.
For now we focus on deterministic impairments and neglect PMD.
We simulate transmission of 112 subcarrier NFDM and OFDM pulses based on a 16QAM constellation  over $N_\text{span}=25$ spans of 80km of standard single-mode fiber.
We assume ideal flat-gain amplifiers with a noise figure of $N_F=6.2$dB. The system parameters are summarized in Table~\ref{tab:params}. They are comparable with those of Ref.~\cite{Le2015} to facilitate a comparison with the single-polarization case.

We first compare single-polarization NFT transmission based on the NLSE with polarization-multiplexed transmission based on the Manakov equation. Figure~\ref{fig:compare_nlse} shows that in the linear regime the two curves are offset by 3dB because polarization multiplexing doubles the transmission power without increasing the signal to noise ratio.
We note that in this calculation we increased the guard band and thereby the sampling rate in frequency domain by a factor of two in the computation of the PDM-NFT (the guard band in transmission is the same) compared to the single-polarization case. This is to avoid penalties due to inaccuracies in the NFT, which here amount to approximately 1dB at peak power if the same guard bands are used. Without this penalty, the $Q$-factor at peak power is roughly the same in both cases. This implies that data rates can approximately be doubled using polarization multiplexing. We show in Sec.~\ref{sec:pmd} that this conclusion still holds in presence of polarization effects.

\begin{figure}%
\begin{center}
\includegraphics[width=0.7\columnwidth]{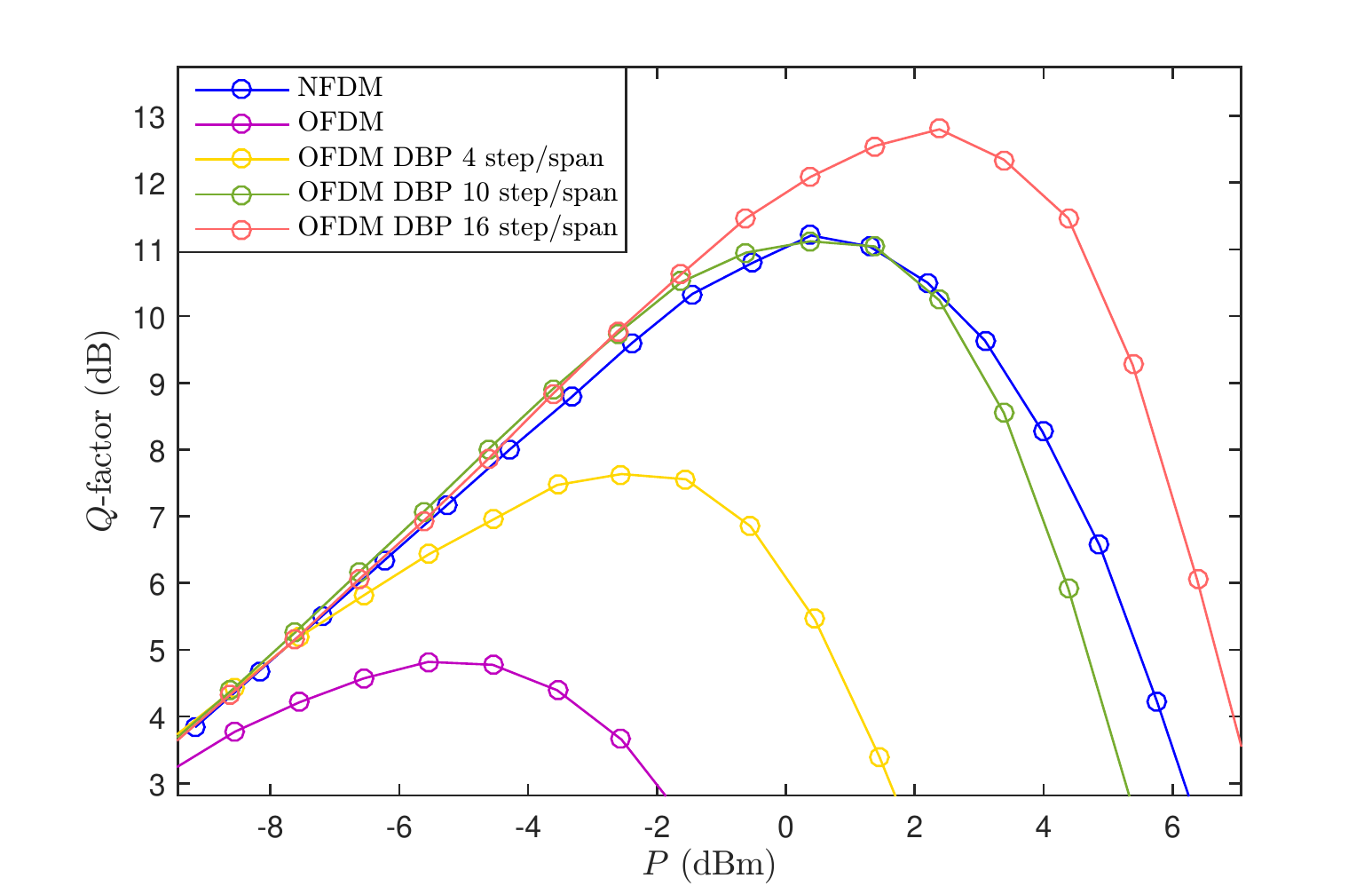}%
\caption{Comparison of polarization-multiplexed OFDM to NFDM transmission in an idealized setting. We use the Manakov equation without PMD-effects for 25 spans of 80 km for 16QAM. NFDM performs as well as OFDM with 10step/span DBP. The full potential of NFDM is leveraged in a multi-channel scenario, where DBP is less efficient.}%
\label{fig:mainresult}%
\end{center}
\end{figure}

\begin{figure}%
\begin{center}
\begin{tabular}{lr}
\includegraphics[width=0.25\columnwidth]{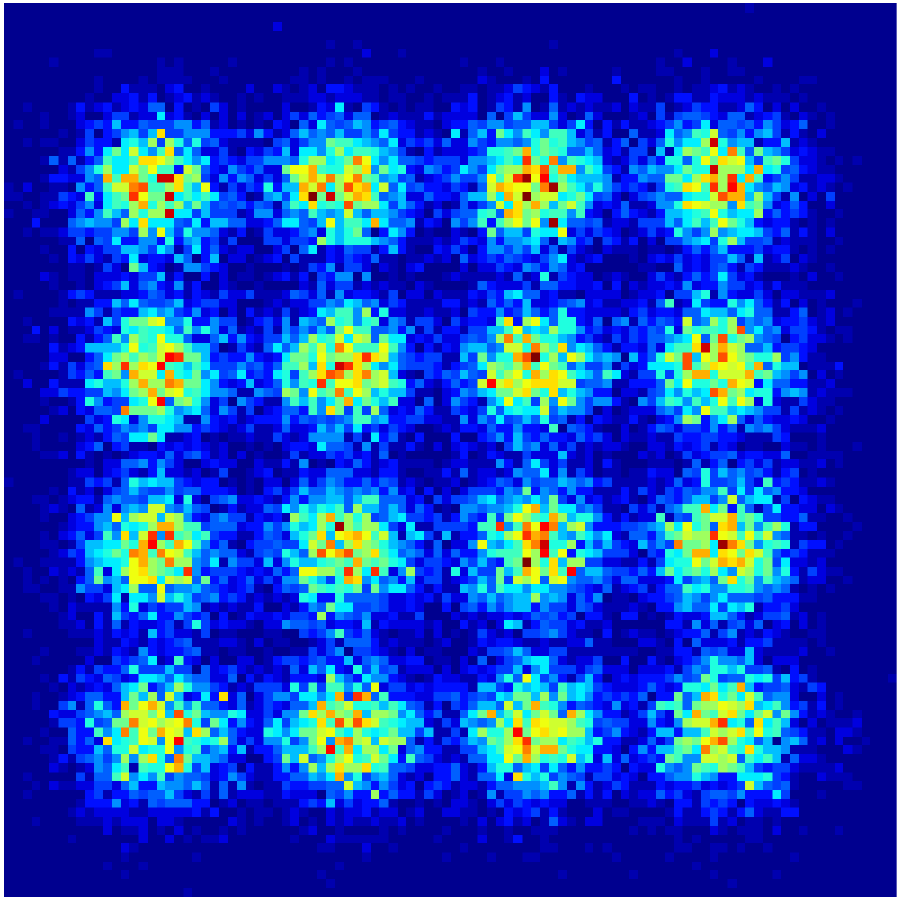}&\includegraphics[width=0.246875\columnwidth]{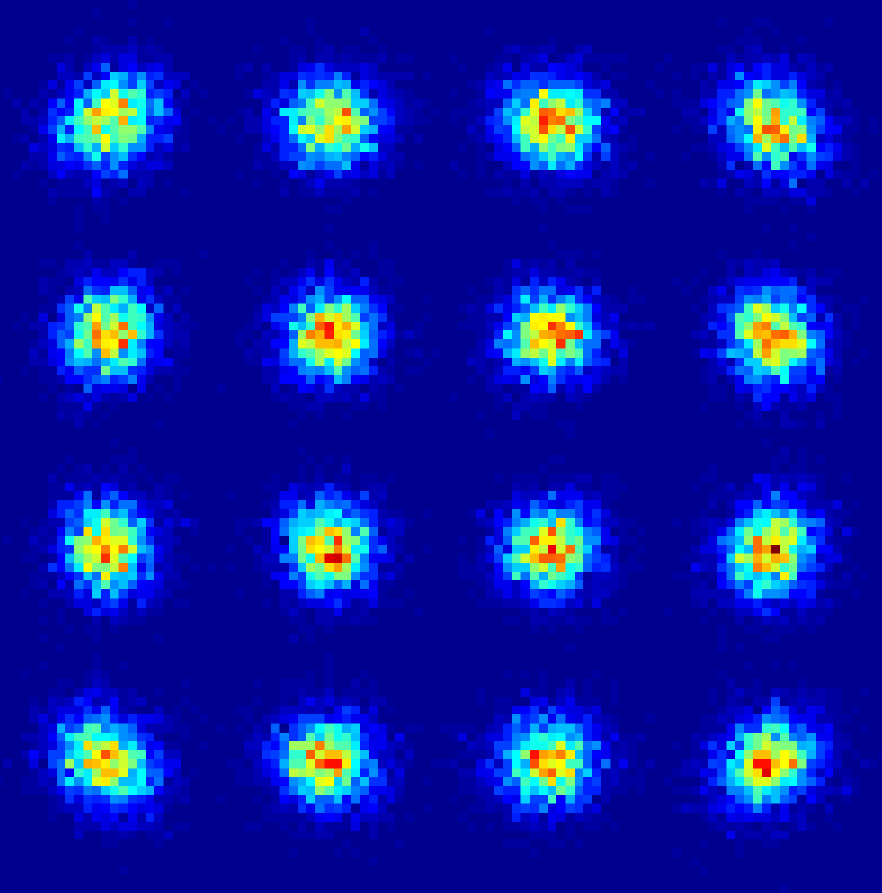}
\end{tabular}
\end{center}
\caption{Received constellations for OFDM (left panel, without DBP) and NFDM (right panel) at the respective optimal launch power. The noise in NFDM is clearly non-Gaussian.}
\label{fig:constellations}
\end{figure}
  
In Fig.~\ref{fig:mainresult} we compare the $Q$-factor as a function of launch power in PDM-OFDM and PDM-NFDM transmission. The optimum launch power in OFDM is significantly smaller than in NFDM. OFDM transmission is limited by nonlinearity, while NFDM is mainly limited by nonlinear signal-noise interaction. 
The $Q$-factor at optimum launch power is significantly higher in NFDM than in OFDM. We observe an overall gain of $6.4$dB in $Q$-factor. Figure~\ref{fig:constellations} shows the corresponding received constellations at the respective optimal launch power.

In Fig.~\ref{fig:mainresult} we further report results obtained by digital backpropagation after OFDM transmission. For a fair comparison with NFDM, we use the same sampling rate in the DBP and apply it without prior downsampling. OFDM with DBP achieves similar performance to NFDM for around 10 DBP steps per span and exceeds it for 16 steps per span. We note that here we considered a single-user scenario. The full potential of NFDM is leveraged in a network scenario, where DBP is less efficient.
Our results are in good qualitative agreement with those reported in Ref.~\cite{Le2015} for the single-polarization case, while achieving twice the data rate.

\subsection{Effect of PMD}
\label{sec:pmd}

\subsubsection{Simulation of PMD}
\label{sec:ssfm}

The fiber model~\eqref{eq:manakov-pmd} describes light propagation in linearly birefringent fibers~\cite{Wai1991,Evangelides1992}.
Here $\Delta\beta_1$ is the difference in propagation constants for the two polarization  states which are induced by fiber imperfections or stress. In real fibers, this so-called modal birefringence varies randomly, resulting in PMD.

We simulate PMD with the coarse-step method, in which continuous variations of the birefringence are approximated by a large number of short fiber sections in which the birefringence is kept constant. PMD is thus emulated in a distributed fashion~\cite{Bertolini2010}.
We use fixed-length sections of length 1km, larger than typical fiber correlation lengths. At the beginning of each section, the polarization is randomly rotated to a new point on the Poincar\'e sphere. We apply a uniform random phase in the new reference frame. The latter accounts for the fact that in reality the birefringence will vary in the sections where it is assumed constant, which will lead to a random phase relationship between the two polarization components~\cite{Wai1991}. The differential group delay (DGD) of each section is selected randomly from a Gaussian distribution. This way artifacts in the wavelength domain caused by a fixed delay for all sections are avoided~\cite{Eberhard2005}.

Within each scattering section, the equation~\eqref{eq:manakov-pmd} without PMD terms is solved using standard split-step Fourier integration. 
To speed up the simulations, we employ a CUDA/C++ based implementation with a MEX interface to Matlab.

The resulting DGD of the fiber is Maxwell distributed~\cite{Poole1991},
\begin{equation}
p_{\av{\Delta t}}(\Delta t) = \frac{32}{\pi^2}\frac{\Delta t^2}{\av{\Delta t}^3} \exp\left(-\frac{4\Delta t^2}{\pi\av{\Delta t}^2}\right).
\label{eq:maxwell}
\end{equation}
For the Maxwell distribution, the mean is related to the root-mean-square (RMS) delay by $\av{\Delta t}\sqrt{3\pi/8}=\sqrt{\av{\Delta t^2}}$. The average DGD varies with the square root of the fiber length~\cite{Agrawal2007},
\begin{equation}
\av{\Delta t}\sim \sqrt{\av{\Delta t^2}} = D_\text{PMD}\sqrt{L},
\end{equation}
where $D_\text{PMD}$ is the PMD parameter.
Typical PMD values for fibers used in telecommunications range from 0.05 in modern fibers to 0.5 ps/$\sqrt{\text{km}}$.

\subsubsection{PMD impact on NFDM}

\begin{figure}%
\begin{center}
\includegraphics[width=0.6\columnwidth]{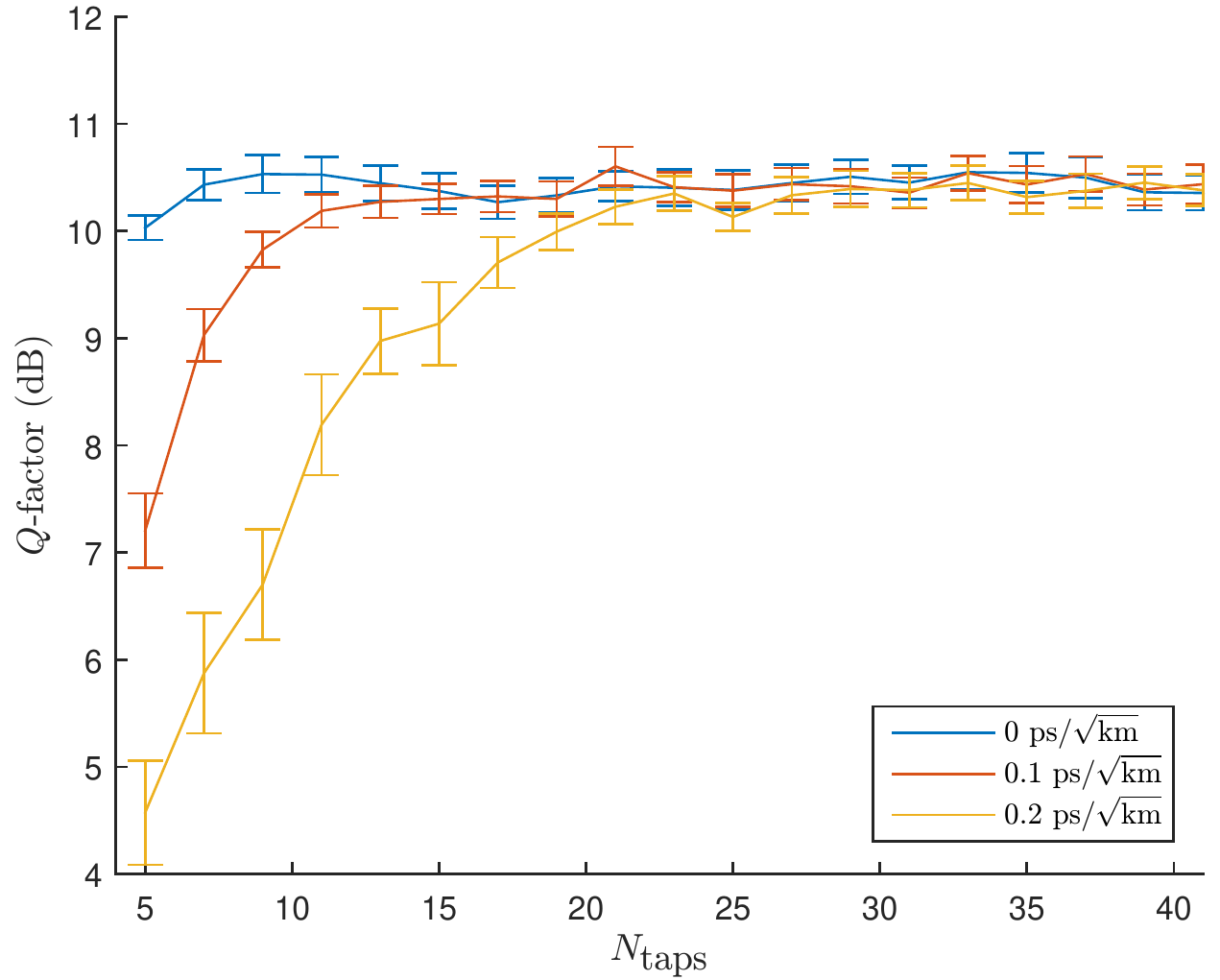}%
\caption{The effect of the number of taps on the Q-factor for different values of $D_\text{PMD}$. The power of the signal in the simulation is 0.5 dBm.}%
\label{fig:pmdtaps}%
\end{center}
\end{figure}

In this section, we consider polarization effects on the NFT transmission. 
Since the nonlinear term in the Manakov equation is invariant under polarization rotations, the equalization can be done in in the nonlinear Fourier domain (cf. Fig.~\ref{fig:system}).
We use a simple training sequence based equalization algorithm to compensate linear polarization effects.
The samples of the first NFDM symbol are used as the training sequence. The filter taps of the equalizer are determined using least-squares estimation.
To obtain the $Q$-factor we average the BER over 120 random realizations of PMD for each data point.

Due to its statistical nature, the effects of PMD are often quantified in terms of outage probabilities. In practice, the system is designed to tolerate a certain amount of PMD, in this case by fixing the number of taps in the equalizer. When the DGD exceeds this margin, the system is said to be in outage. 
We first determine the number of taps to achieve a given outage probability. Fig.~\ref{fig:pmdtaps} shows the $Q$-factor as a function of the number of taps for different values of the PMD parameter. For $D_\text{PMD}=0$ we only have random polarization rotations and no DGD. Hence there is no interaction between nonlinearity and PMD and we use this case as the reference. The upturn for a small number of taps is due to the fact that a couple of taps are required to fully compensate the polarization rotations in presence of noise. 

For finite PMD, the curves for different PMD values converge to the same result within error bars. We can estimate the required number of taps from the Maxwell distribution. In order for the system to fully compensate PMD in 98.7\% of the cases (1.3\% outage probability), the equalization must cover a time interval equal to the mean plus 3 standard deviations of the DGD distribution. For example, for $D_\text{PMD} = 0.1$ps$/\sqrt{\text{km}}$ this interval corresponds to $12.8$ps. Using the sampling frequency we find that we approximately need 12 taps to converge, consistent with the figure.

We note that by construction our equalizer also corrects potential phase rotations and therefore at least part of the nonlinear effects. In order to separate effects due to interaction of nonlinearity and PMD, we performed a calculation where we reversed the polarization rotations and phases exactly (instead of using the equalizer), by keeping track of the randomly generated angles and DGD values in the fiber simulation. We find a small phase rotation consistent with but smaller than the one reported in~\cite{Tavakkolnia2017a}. In our case it remains negligible at peak power.
\begin{figure}%
\begin{center}
\includegraphics[width=0.6\columnwidth]{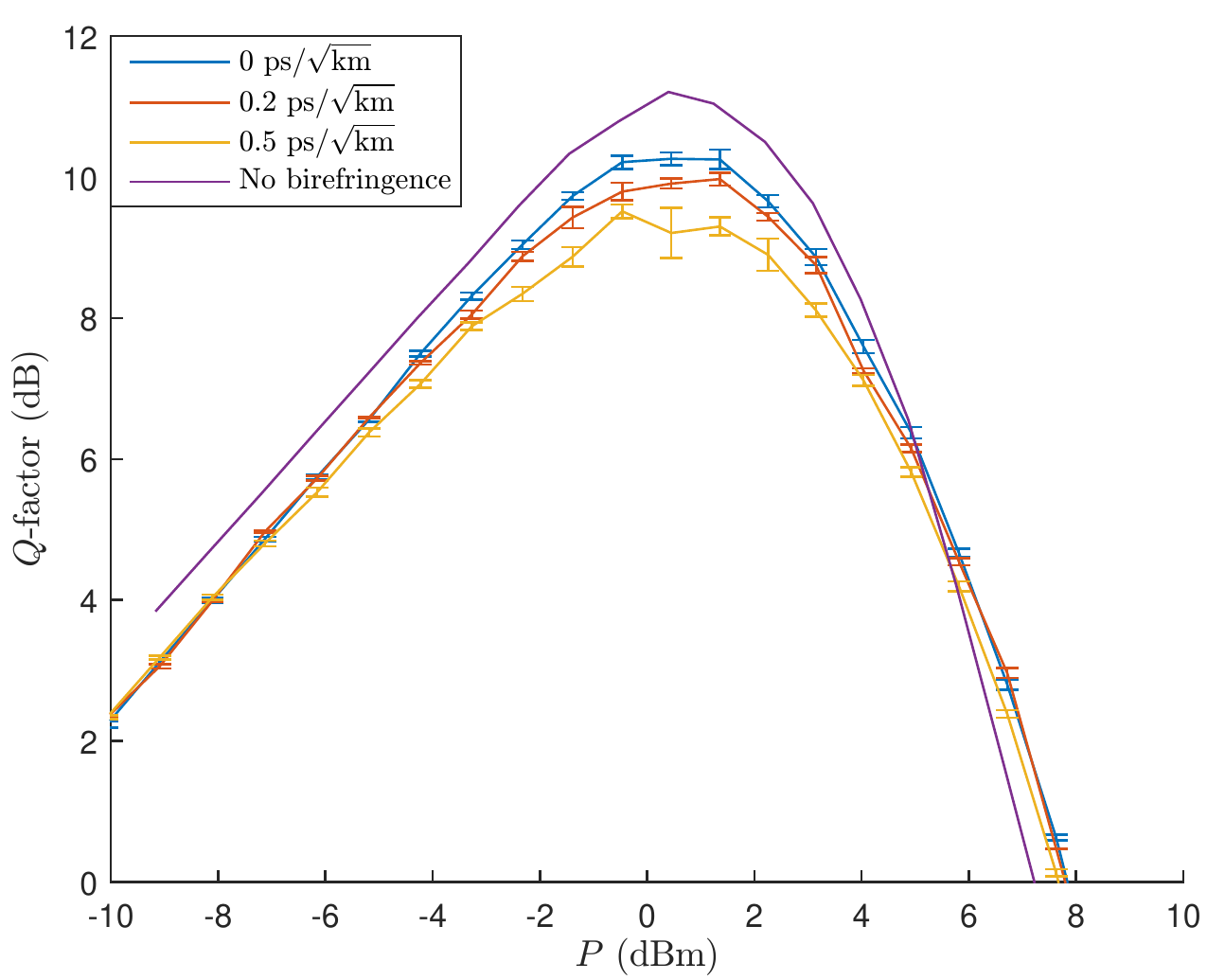}%
\caption{The effect of PMD on PDM-NFDM. We used 13, 25 and 61 taps in the equalizer for $D_\text{PMD}=0,0.2$ and $0.5$ps$/\sqrt{\text{km}}$, respectively.}%
\label{fig:pmdpower}%
\end{center}
\end{figure}

In Fig.~\ref{fig:pmdpower} we compare the $Q$-factor as a function of launch power with and without PMD effects. Here we use the number of taps determined as described above.
We observe a penalty of roughly 0.3dB at peak power for $D_\text{PMD}=0.2$ps$/\sqrt{\text{km}}$ relative to the case without PMD ($D_\text{PMD}=0$ps$/\sqrt{\text{km}}$). The penalty is therefore not serious for typical fibers used in telecommunication.
Compared to the case without PMD effects and random birefringence (labeled ``no birefringence"), we find a penalty of roughly 1.2dB at peak power for the case of zero PMD due to the equalization.

\section{Conclusions}

In this paper, we have proposed polarization-division multiplexing based on the nonlinear Fourier transform. NFT algorithms are developed based on the Manakov equations. Our simulations demonstrate feasibility of polarization multiplexed NFDM transmission over standard single-mode fiber. The results show that data rates can approximately be doubled in polarization-multiplexed transmission compared to the single-polarization case. This is an important step to achieve data rates that can exceed those of conventional linear technology.

Numerical simulations of polarization multiplexed transmission over a realistic fiber model including randomly varying birefringence and polarization mode dispersion have shown that penalties due to PMD in real fibers do not seriously impact system performance.
Our fiber simulations are based on the Manakov equations. As a next step, which is beyond the scope of this paper, the results should be verified experimentally.

The equation governing light propagation of $N$ modes in multi-mode fibers in the strong coupling regime can be written in the form~\cite{Mumtaz2013}
\begin{align}
\frac{\partial \vc{A}}{\partial Z} =& j\frac{\bar{\beta}_2}{2}\frac{\partial^2 \vc{A}}{\partial T^2} 
- j\gamma\kappa\norm{\vc{A}}^2\vc{A},
\label{eq:multimode}
\end{align}
where $\vc A=(A_1,\dots,A_N)^T$ and $\bar{\beta}_2$ denotes the average group velocity dispersion.
The NFT, as well as the algorithms presented here, generalize to this equation in a straightforward manner. Our approach therefore paves the way to combining the nonlinear Fourier transform with space-division multiplexing.

\section{Acknowledgements}
J-W.G. would like to thank Wasyhun A. Gemechu for helpful discussions.
H.H. acknowledges useful discussions with Djalal Bendimerad, Majid Safari, Sergei K. Turitsyn and Huijian Zhang.

\end{document}